\documentclass[12pt,epsf]{article}
\usepackage{graphics}
\usepackage{graphicx}

\input{epsf}
\begin{document}
\begin{titlepage}
\begin{center}

 \vspace{-0.1in}

{\large \bf The Bekenstein Bound \\ in \\ Asymptotically Free Field Theory }\\
\vspace{.5in}{\large\em E. Arias\,\,\footnotemark[1],
 N. F. Svaiter\,\,\footnotemark[2]}\\
\vspace{.1in}

Centro Brasileiro de Pesquisas F\'{i}sicas-CBPF\\
Rua Dr. Xavier Sigaud 150,
Rio de Janeiro, RJ,\,22290-180, Brazil\\

\vspace{.3in}{\large\em G. Menezes \,\footnotemark[3]}\\

\vspace{.1in}
Instituto de F\'{\i}sica Te\'{o}rica, Universidade Estadual Paulista,\\
 Rua Dr. Bento Teobaldo Ferraz 271, Bloco II, Barra Funda,\\
 S\~ao Paulo, SP,  01140-070, Brazil \\

\subsection*{\\Abstract}
\end{center}

\baselineskip .10in

For spatially bounded free fields, the Bekenstein bound states
that the specific entropy satisfies the inequality $\frac{S}{E}
\leq 2 \pi R$, where $R$ stands for the radius of the smallest
sphere that circumscribes the system. The validity of the
Bekenstein in the asymptotically free side of the Euclidean
$(\lambda\,\varphi^{\,4})_{d}$ scalar field theory is
investigated. We consider the system in thermal equilibrium with a
reservoir at temperature $\beta^{\,-1}$ and defined in a compact
spatial region without boundaries. Using the effective potential,
we discuss the thermodynamic of the model. For low and high
temperatures the system presents a condensate. We present the
renormalized mean energy $E$ and entropy $S$ for the system and
show in which situations the specific entropy satisfies the
quantum bound.
\vspace{0,3cm}\\
PACS numbers: 03.70+k, 04.62.+v

\footnotetext[1]{e-mail:\,\,enrike@cbpf.br}
\footnotetext[2]{e-mail:\,\,nfuxsvai@cbpf.br}
\footnotetext[3]{e-mail:\,\,gsm@ift.unesp.br}

\end{titlepage}
\newpage\baselineskip .18in

\section{Introduction}
\quad \,\, There have been a lot of activities discussing
classical and quantum fields in the presence of macroscopic
boundaries. These subjects raise many interesting questions. A
basic question that has been discussed in this scenario, when
quantum fields interact with boundaries, is about the issue that
these systems may be subjected to certain fundamental bounds. 't
Hooft \cite{hooft} and Susskind \cite{suss}, combining quantum
mechanics and gravity, introduced the holographic entropy bound
$S\leq \pi\,c^{3}\,R^{2}/\hbar\,G$ \cite{bousso}. This holographic
bound relates information not with the volume, but with the area
of surfaces. Another bound is the Bekenstein bound, which relates
the entropy $S$ and the mean energy $E$ of a quantum system with
the size of the boundaries that confine the fields. It is given by
$S \leq 2 \pi\,E\, R/\hbar\,c$, where $R$ stands for the radius of
the smallest sphere that circumscribes the system \cite{bek0}
\cite{bek1} \cite{bek4} \cite{bek2} \cite{beki5}.

The aim of this paper is to investigate the validity of the
Bekenstein bound in systems defined in a compact spatial region
without boundaries, described by asymptotically free theories. We
study the ordinary Euclidean $(\lambda\varphi^{\,4})_{d}$ massless
scalar field theory, with a negative sign of the coupling constant
\cite{prova} \cite{bender2000} \cite{bender2001}. This field
theory is renormalizable in a four-dimensional space-time,
asymptotically free and has a nontrivial vacuum expectation value.

Studying the $(\lambda\,\varphi^{\,p})_{d}$ self-interacting
massless scalar field theory in the strong-coupling regime at
finite temperature, and also assuming that the field is confined
in a compact spatial region, a generalization for the Bekenstein
bound was obtained by Aparicio Alcalde \textit{et al}
\cite{bound8}. The basic problem that arises in theories with
non-linear fields is the possibility of non-linear interactions
change the energy spectrum of the system invalidating the quantum
bound. Previous works studying the bound in weakly coupled fields
can be found in Refs. \cite{bek3} \cite{int}. Bekenstein and
Guedelman studied the massless charged self-interacting scalar
field in a box and proved that in this case non-linearity does not
violate the bound on the specific entropy. In Ref. \cite{bound8}
it was assumed that the fields are defined in a simply connected
bounded region, i.e., a hypercube of size $L$, where the scalar
field satisfies Dirichlet boundary conditions. Working in the
strong-coupling regime of the $(\lambda\,\varphi^{\,p})_{d}$ field
theory and making use of the strong coupling expansion
\cite{kovesi} \cite{meni} \cite{be1} \cite{kla22} \cite{novo1}, it
was obtained the renormalized mean energy and the entropy for the
system up to the order $\lambda^{-\frac{2}{p}}$, presenting an
analytic proof that the specific entropy also satisfies in some
situations a quantum bound. Considering the low temperature
behavior of the thermodynamic quantities of the system, it was
shown that for negative renormalized zero-point energy, the
quantum bound can be invalidated. Note that a still open question
is how the sign of the renormalized zero-point energy of free
fields described by Gaussian functional integrals depends on the
topology, dimensionality of the spacetime, the shape of bounding
geometry or others physical properties of the system
\cite{ambjorn} \cite{caruso} \cite{robson} \cite{amaral}. For
complete reviews discussing the Casimir effect \cite{casimir}, see
for example the Refs. \cite{plunien} \cite{mamayev} \cite{krech}
\cite{bordag} \cite{milton}.

The purpose of this article is to investigate another physical
situation that has not been discussed in the literature. We should
note that a step that still remains to be derived is the validity
of the bound for the case of interacting field theory described by
asymptotically free models \cite{ass1} \cite{ass2} \cite{ass3}
\cite{ass4} \cite{ass5}, at least up to some order of perturbation
theory. This situation of a deconfined field theory with
asymptotically free behavior, defined in a small compact region of
space may occur in $QCD$ in the confinement-deconfinement phase
transition at high temperatures or if usual matter is strongly
compressed \cite{shuryak} \cite{gross} \cite{mclerran}. For a
complete review of the subject see Ref. \cite{livroshuryak}. In
ultra-relativistic heavy ion collisions we expect that the plasma
of quarks and gluons can be produced, just after the collision,
hot and compressed nuclear matter is confined in a small region of
space. No practical method had been developed to solve $QCD$, and
therefore the basic problem we have discussed it remains
unanswered, unless we try to describe a simpler model which
develop asymptotic freedom for some values of the coupling
constant.

In order to investigate the Bekenstein bound in this
asymptotically free theory, we assume that the scalar field is
confined in a bounded region. Working in the weak-coupling
perturbative expansion with the $(\lambda\,\varphi^{\,4})_{d}$ we
assume periodic boundary conditions in all spatial directions, in
order to maintain translational invariance of the model. This same
approach was used in Ref. \cite{ford}. For papers studying
non-translational invariant systems and analyzing the divergences
of the theory see, for example, the Refs. \cite{sy} \cite{fosco}
\cite{caicedo} \cite{nfs} \cite{robson1} \cite{robson2}
\cite{aparicio}. We also assume that the system is in thermal
equilibrium with a reservoir and investigate the asymptotic free
side of the $(\lambda\,\varphi^{\,4})_{d}$ \cite {prova}
\cite{brandt1} \cite{brandt2} \cite{riva} \cite{gaw} \cite{lang}.
In order to study the existence of a quantum bound on the specific
entropy, we study the behavior of the specific entropy using the
effective action method.

We would like to point out that the theory with a negative
coupling constant develops a condensate as was shown by Parisi
\cite{parisi}. In the self-interaction
$(\lambda\,\varphi^{\,4})_{d}$ field theory, it is possible to
find the vacuum energy $E(\lambda)$. This quantity is given by the
sum of all vacuum-to-vacuum connected diagrams. In the $\lambda$
complex plane, the function $E(\lambda)$ is analytic for
$Re(\lambda)>0$ and the discontinuity on the negative real axis is
related to the mean life of the vacuum. For a system with $N$
particles, let us define $E_N(\lambda)$ as the energy of such
state. For negative $\lambda$, there are collapsed states of
negative energy. Defining $\max E_N(\lambda) = E_B$, the
probability of the vacuum to decay is $e^{-E_B}$. The particles on
the collapsed state will be described by a classical field
$\varphi_{_{0}}$.

The organization of the paper is as follow: in Sec. II we study
the effective potential of the theory at the one-loop level. Due
to the boundary conditions imposed on the field, there is a
topological generation of mass. The topological squared mass
depends on the ratio $\xi=\beta/L$, and its sign is critical to
the profile of the effective potential. In Sec. III we present our
results of the thermodynamic functions and study the validity of
the Bekenstein bound in the model. To simplify the calculations we
assume the units to be such that $\hbar=c=k_{B}=1$.

\section{The effective potential at the one loop level}\

Let us consider a neutral scalar field with a
$(\lambda\varphi^{4})$ self-interaction, defined in a
$d$-dimensional Minkowski spacetime. The generating functional of
all vacuum expectation value of time-ordered products of the
theory has a Euclidean counterpart, that is the generating
functional of complete Schwinger functions. The
$(\lambda\varphi^{4})_{d}$ Euclidean theory is defined by these
Euclidean Green's functions. The Euclidean generating functional
$Z(h)$ is defined by the following functional integral
\cite{martin1} \cite{livron}:
\begin{equation}
Z(h)=\int [d\varphi]\,\, \exp\left(-S_{0}-S_{I}+ \int d^{d}x\,
h(x)\varphi(x)\right), \label{1}
\end{equation}
where the action that describes a free scalar field is given by
\begin{equation}
S_{0}(\varphi)=\int d^{d}x\,
\left(\frac{1}{2}(\partial\varphi)^{2}+\frac{1}{2}
m_{0}^{2}\,\varphi^{2}\right). \label{2}
\end{equation}
The interacting part, defined by the non-Gaussian contribution, is
given by the following term in the action:
\begin{equation}
S_{I}(\varphi)= \int d^{d}x\,\frac{\lambda}{4!} \,\varphi^{4}(x).
\label{3}
\end{equation}
In Eq. (\ref{1}), $[d\varphi]$ is formally given by
$[d\varphi]=\prod_{x} d\varphi(x)$, $m_{0}^{2}$ and $\lambda$ are
the bare squared mass and coupling constant respectively. Finally,
$h(x)$ is a smooth function that is introduced to generate the
Schwinger functions of the theory by functional derivatives.

We are assuming a spatially bounded system in equilibrium with a
thermal reservoir at temperature $\beta^{-1}$. Assuming that the
coupling constant is a small parameter, the weak-coupling
expansion can be used to compute the partition function defined by
$Z(\beta,\Omega,h)|_{\,h=0}$, where $h$ is a external source and
we are defining the volume of the $(d-1)$ manifold as
$V_{d-1}\equiv\,\Omega$. From the partition function we define the
free energy of the system, given by
$F(\beta,\Omega)=-\frac{1}{\beta}\ln\,Z(\beta,\Omega,h)|_{\,h=0}$.
This quantity can be used to derive the mean energy
$E(\beta,\Omega)$, defined as
\begin{equation}
E (\beta,\Omega) = - \frac{\partial }{\partial\beta}\ln
Z(\beta,\Omega,h)|_{\,h=0}, \label{imp}
\end{equation}
and the canonical entropy $S(\beta,\Omega)$ of the system is given
by
\begin{equation}
S (\beta,\Omega)= \biggl(1 - \beta \frac{\partial}{\partial\beta}
\biggr)\ln Z(\beta,\Omega,h)|_{h=0}. \label{imp1}
\end{equation}

Since the scalar theory with the negative coupling constant
develops a condensate, it is convenient to work with the effective
potential of the system. As was stressed by Bender et. al.
\cite{bender2000}, non-perturbative techniques must be used to
find the true vacuum of the system. Therefore let us study first
the effective potential at the one-loop level associated to a
self-interacting scalar field defined in a $d$-dimensional
Euclidean space.

Let us consider that the system is in thermal equilibrium with a
reservoir at temperature $\beta^{\,-1}$. Therefore we assume the
Kubo-Martin-Schwinger (KMS) condition \cite{matsu} \cite{kubo}
\cite{martin} \cite{kapusta}. We will work with a massless scalar
field and assume $d=4$, and in order to simplify the calculations
we impose periodic boundary conditions for the field in all three
spatial directions, with compactified lengths $L_1$,$L_2$ and
$L_3$.  The Euclidean effective potential can be written as:
\begin{eqnarray}
V_{eff}(\phi\,;\beta,L_1,L_2,L_3)&=&\frac{\mu^4}{3}
\pi^2g\phi^4+U+counterterms+
\nonumber\\
&&\frac{1}
{\beta\Omega}\sum_{s=1}^{\infty}\frac{(-1)^{s+1}}{2s}g^s\phi^
{2s}Z_4(2s,a_1,a_2,a_3,a_4), \label{h1}
\end{eqnarray}
where we have defined the quantities $\phi=\varphi/\mu$,
$g=\lambda/8\pi^2$, $a_i^{-1}=\mu L_i$ $(i=1,2,3)$,
$a_4^{-1}=\mu\beta$, $\Omega=L_1L_2L_3$ and finally $Z_4(2s,
a_1,...,a_4)$ is the Epstein zeta function \cite{ambjorn}. Note
that we have introduced a mass parameter $\mu$ in order to keep
the Epstein zeta function, $Z_4$, a dimensionless quantity.

The first contribution to the effective potential given in Eq.
(\ref{h1}) is the classical potential. The second contribution
$U(\beta,L_{1},L_{2},L_{3})$, is given by
\begin{equation}
U(\beta,L_{1},L_{2},L_{3})=\frac{1}{2\beta\Omega}
\sum_{n_1,...,n_4=-\infty}^{\infty\hspace{0.18cm},}
\ln\left(\left(\frac{2\pi
n_1}{L_1}\right)^2+ \left(\frac{2\pi
n_2}{L_2}\right)^2+\left(\frac{2\pi n_3}{L_3}\right)^2+
\left(\frac{2\pi n_4}{\beta}\right)^2\right).
\label{h2}
\end{equation}
The prime that appears in the Eq. (\ref{h2}) indicates that the
term for which all $n_i=0$ must be omitted.  We can rewrite Eq.
(\ref{h2}) as
\begin{equation}
U(\beta,L_{1},L_{2},L_{3})=\frac{1}{\beta\Omega}
\sum_{n_1,...,n_3=-\infty}^{\infty\hspace{0.18cm},}
\biggl(\pi\beta\bar{n}+\ln\left(1-e^{-2\pi\beta\bar{n}}\right)\biggr)+
\frac{1}{\beta\Omega}J_1, \label{h3}
\end{equation}
where we are defining the quantity $\bar{n}(L_{1},L_{2},L_{3})$
and the (infinite) constant $J_1$ as
\begin{equation}
\bar{n}=\sqrt{\Biggl(\frac{n_1}{L_1}\Biggr)^2+\Biggl
(\frac{n_2}{L_2}\Biggr)^2+\Biggl(\frac{n_3}{L_3}\Biggr)^2}
\label{h4}
\end{equation}
and
\begin{equation}
J_1=\sum_{n_1,...,n_3=-\infty}^{\infty\hspace{0.18cm},}
\sum_{m=-\infty}^{\infty\hspace{0.18cm},}\ln\biggl(1+(2\pi
m)^2\biggr)-
\sum_{n_1,...,n_3=-\infty}^{\infty\hspace{0.18cm},}\biggl(1+2\ln(1-e^{-1})\biggr).
\label{h5}
\end{equation}
The last term of the Eq. (\ref{h1}) is explicitly the one-loop
correction to the effective potential, defined in terms of the
homogeneous Epstein zeta function $Z_{p}(2s,a_1,...,a_p)$ given in
Ref. \cite{ambjorn} by
\begin{equation}
Z_{p}(2s,a_1,...,a_p)=\sum_{n_1,...,n_p=-\infty}^{\infty\hspace{0.18cm},}
\left((a_1 n_1)^2+...+(a_p n_p)^2\right)^{-s}.
\label{h6}
\end{equation}
The summation give by Eq. (\ref{h6}) is convergent for $s>p/2$.
The homogeneous Epstein zeta function has an analytic extension to
the complex plane $s\in C$, except for a pole in $s=p/2$. Since
the unique polar contribution occurs for the case in $s=2$, the
theory can be renormalized using only a unique counterterm,
introduced to renormalize the coupling constant of the theory.
Since we are assuming periodic boundary conditions for the field
in all spatial directions, it appears a topological generation of
mass, coming from the self-energy Feynman diagram
\cite{fordyoshimura} \cite{fordtop} \cite{fordbir}. The
topological mass is defined in terms of the first renormalization
condition given by
\begin{equation}
\frac{\partial^2 V_{eff}}{\partial\phi^2}\bigg|_{\phi=0}=m_T^2\mu^2.
 \label{h7}
\end{equation}
Using the Epstein zeta function, the topological squared mass
$m_T^2$ can be written as
\begin{equation}
m_T^2=\frac{g}{\mu^2\beta\Omega}Z_4(2,a_1,a_2,a_3,a_4).
 \label{h8}
\end{equation}
The above result was obtained also by Elizalde and Kirsten
\cite{eliki}. As was discussed by these authors, the topological
squared mass depends on the values of the compactified lengths and
the temperature. For simplicity we will call this quantity a
topological mass. The next step is to study the two cases
$m_T^2>0$ and $m_T^2<0$ separately.

\subsection{The positive topological squared mass, i.e., $m_T^2>0$}

\begin{figure}
\centering
\includegraphics{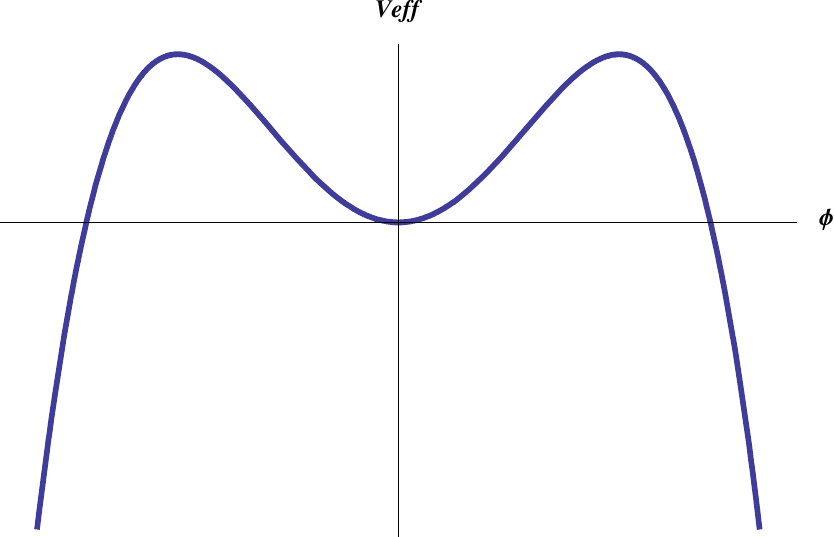}
\caption{The effective potential for the case where $m_T^2>0$}
\label{figure1}
\end{figure}

First, let us write the effective potential in the form
\begin{eqnarray}
V_{eff}(\phi;\beta,L_1,L_2,L_3)&=&\mu^2\frac{m_T^2}{2}\phi^2+
\frac{\mu^4}{3}\pi^2g\phi^4+\mu^4\frac{\delta\lambda}{4!}\phi^4+
U+\nonumber\\&&\frac{1}{\beta\Omega}\sum_{s=2}^{\infty}\frac{(-1)^{s+1}}
{2s}g^s\phi^{2s}Z_4(2s,a_1,a_2,a_3,a_4).
 \label{h9}
\end{eqnarray}
We begin studying the case $m_T^2>0$. We will consider first
particular values of the compactified lengths and temperature in
such a way that the analytic extension of the homogeneous Epstein
zeta function $Z_4(2,a_1,...,a_4)$ takes only negative. Therefore
we consider that the coupling constant is negative, i.e.,
$g=-|g|<0$. In this case we will have that the topological squared
mass is given by
\begin{equation}
m_T^2=-\frac{|g|}{\mu^2\beta\Omega}Z_4(2,a_1,a_2,a_3,a_4).
 \label{h10}
\end{equation}
Therefore we get a physical mass of a scalar particle confined
inside our finite domain. The second renormalization condition,
which gives a finite coupling constant is
\begin{equation}
\frac{\partial^4 V_{eff}}{\partial\phi^4}\bigg|_{\phi=0}=8\pi^2 g\mu^4.
 \label{h11}
\end{equation}
Using Eq. (\ref{h11}) in Eq. (\ref{h9}) we can find the
renormalized effective potential. In this case (for negative
coupling constant, $Z_4(2,a_1,...,a_4)$ taking only negative
values and hence $m_T^2>0$) it can be written as
\begin{eqnarray}
V_{eff}^R(\phi;\beta,L_1,L_2,L_3)&=&\mu^2\frac{m_T^2}{2}\phi^2-\frac{\mu^4}
{3}\pi^2|g|\phi^4+U\nonumber\\&&-\frac{1}{\beta\Omega}\sum_{s=3}^{\infty}
\frac{|g|^s}{2s}\phi^{2s}Z_4(2s,a_1,a_2,a_3,a_4).
 \label{h12}
\end{eqnarray}
The renormalized effective potential is presented in Fig.
(\ref{figure1}).
It has a local metastable minimum at the origin and it is not
bounded from below. This is an expected result since the model is
the asymptotically free side of the Euclidean
$(\lambda\,\varphi^{\,4})_{d}$ scalar field theory.

\begin{figure}
 \centering
 \includegraphics{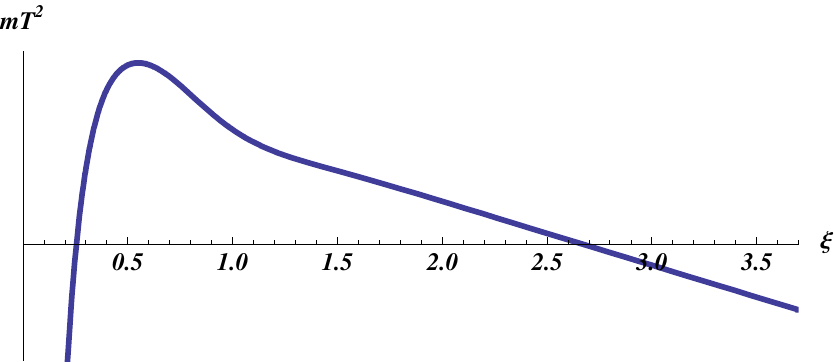}
 \caption{Behavior of the $m_T^2$ with $\xi$}
\label{figure2}
\end{figure}

Next, let us calculate the specific entropy $S/E$, where the mean
energy $E$ and the entropy $S$ are given by Eq. (\ref{imp}) and
Eq. (\ref{imp1}). First we should perform and inverse Legendre
transform in order to obtain $\ln Z(\beta,\Omega,h)$. Note that
these thermodynamics functions are calculated in the absence of
the source, i.e., $h=0$. In terms of the effective potential, we
have to find the stationary point of the renormalized effective
potential, $\phi_{\,0}$, defined by the equation
\begin{equation}
\frac{\partial V_{eff}^R}{\partial\phi}\bigg|_{\phi=\phi_0}=0.
 \label{h13}
\end{equation}
Substituting the Eq. (\ref{h12}) in the Eq. (\ref{h13}) we obtain
that $\phi_0$ must satisfies
\begin{equation}
\mu^2m_T^2\phi_0-\frac{4}{3}\mu^4\pi^2|g|\phi_0^3-\frac{1}
{\beta\Omega}\sum_{s=3}^{\infty}|g|^s\phi_0^{2s-1}Z_4(2s,a_1,a_2,a_3,a_4)=0.
 \label{h14}
\end{equation}
From Fig. (\ref{figure1}) we see that Eq. (\ref{h14}) has three
solutions. Since we are interested in the configuration which is
stable under small external perturbations, we take the solution
with the local minimum of the effective potential, i.e.,
$\phi_0=0$. Performing the Legendre transform when the effective
potential reaches its metastable stationary point, we get that
$\ln Z(\beta,\Omega)$ is given by
\begin{eqnarray}
\ln Z(\beta,\Omega)&&=\ln
Z(\beta,\Omega,h)|_{h=0}\nonumber\\&&=-(\beta\Omega)
V_{eff}^R(\phi\,;\beta,L_1,L_2,L_3)|_{\phi=\phi_0=0}.
 \label{h15}
\end{eqnarray}
Substituting the Eq. (\ref{h3}) and Eq. (\ref{h12}) in Eq.
(\ref{h15}) we get
\begin{equation}
\ln
Z(\beta,\Omega)=-\sum_{n_1,...,n_3=-\infty}^{\infty\hspace{0.18cm},}
\left(\pi\beta\bar{n}+\ln\left(1-e^{-2\pi\beta\bar{n}}
\right)\right)-J_1.
 \label{h16}
\end{equation}
The mean energy $E(\beta,\Omega)$ and the canonical entropy
$S(\beta,\Omega)$ of the system in equilibrium with a reservoir
can be derived using Eq. (\ref{imp}), Eq. (\ref{imp1}) and
Eq. (\ref{h16}). We have
\begin{equation}
E(\beta,\Omega)=\sum_{n_1,...,n_3=-
\infty}^{\infty\hspace{0.18cm},}\left(\bar{n}\pi+\frac{2\bar{n}\pi}
{e^{2\bar{n}\pi\beta}-1}\right)
\label{h17}
\end{equation}
and
\begin{eqnarray}
S(\beta,\Omega)=\sum_{n_1,...,n_3=-\infty}^{\infty\hspace{0.18cm},}
\left(\frac{2\bar{n}\pi\beta}{e^{2\bar{n}\pi\beta}-1}-\ln\left
(1-e^{-2\bar{n}\pi\beta}\right)\right)-J_1.
\label{h18}
\end{eqnarray}
Note that we have an infinite constant in the definition of the
canonical entropy. This ambiguity will be circumvented later using
the third law of thermodynamics, and assuming the continuity of
the entropy. For simplicity we will assume that the lengths of
compactification of the spacial coordinates are all equals
$L_i=L$, for $i=1,2,3$, and we will define the dimensionless
variable $\xi=\beta/L$. In this case we can write the mean energy
and the canonical entropy as:
\begin{equation}
E(\xi)=(\varepsilon^{(r)}+P(\xi))/L
 \label{h22}
\end{equation}
and
\begin{equation}
S(\xi)=\xi P(\xi)+R(\xi)+cte.
 \label{he23}
\end{equation}
In Eq. (\ref{h22}) the quantity $\varepsilon^{(r)}$ is defined
by
\begin{equation}
\varepsilon^{(r)}=\sum_{n_1,...,n_3=-\infty}^{\infty\hspace{0.18cm},}\tilde{n}\pi,
\label{er}
\end{equation}
where the variable $\tilde{n}$ is defined as
$\tilde{n}=\sqrt{(n_1)^2+(n_2)^2+(n_3)^2} $. The term
$\varepsilon^{(r)}$ is just the renormalized Casimir energy of the
massless scalar field where we impose periodic boundary conditions
in the three spatial coordinates. In the Ref. \cite{ambjorn} it
was shown that $\varepsilon^{(r)}=-0.81$. The positive functions
$P(\xi)$ and $R(\xi)$ are defined by
\begin{equation}
P(\xi)=\sum_{n_1,...,n_3=-\infty}^{\infty\hspace{0.18cm},}
\bigg(\frac{2\tilde{n}\pi}{e^{2\tilde{n}\pi\xi}-1}\bigg)
\label{P}
\end{equation}
and
\begin{equation}
R(\xi)=-\sum_{n_1,...,n_3=-\infty}^{\infty\hspace{0.18cm},}\ln\left
(1-e^{-2\tilde{n}\pi\xi}\right).
\label{R}
\end{equation}
Note that the situation where $m_T^2>0$ is satisfied only for some
specific values of the ratio between $\beta$ and $L$, given by
$\xi$. Using the analytic extensions presented in Ref.
\cite{eliki}, we can write the topological squared mass as
\begin{eqnarray}
m_T^2=-\frac{|g|}{L^2}\frac{f_1(\xi)}{\xi},
 \label{e1}
\end{eqnarray}
where the function $f_1(\xi)$ is the analytic extension of
$Z_4(2s,1,1,1,\xi^{-1})$ at $s=1$, and is given by
\begin{equation}
f_1(\xi)=a\xi+\frac{\pi^2}{3}\xi^2+K(\xi).
 \label{e2}
\end{equation}
The coefficient $a$ and the function $K(\xi)$ in Eq. (\ref{e2})
are respectively given by
\begin{eqnarray}
a&=&2\pi\gamma
+2\pi\ln\frac{1}{4\pi}+\frac{\pi^2}{3}+8\pi\sum_{n,n_1=1}^
{\infty}\Big(\frac{n_1}{n}\Big)^{1/2}K_{1/2}(2\pi
nn_1)
\nonumber\\&&+4\pi\xi\sum_{n=1}^{\infty}
\sum_{n_1,n_2,n_3=-\infty}^{\infty\hspace{0.18cm},}K_0\Big(2\pi
n\sqrt{n_1^2+n_2^2}\Big) \label{e3}
\end{eqnarray}
and
\begin{equation}
K(\xi)=4\pi\xi^{3/2}\sum_{n=1}^{\infty}
\sum_{n_1,n_2,n_3=-\infty}^{\infty\hspace{0.18cm},}
\Big(\frac{\tilde{n}}{n}\Big)^{-1/2}K_{-1/2}(2\pi n\tilde{n}\xi)
\label{e4}.
\end{equation}
The functions $K_r(\xi)$ that appears in Eq. (\ref{e3}) and Eq.
(\ref{e4}) are the Kelvin functions \cite{grads}. In Fig.
(\ref{figure2}) the behavior of the topological squared mass is
presented. There are three regions of values of $\xi$ where the
topological squared mass has a defined sign. They are given
respectively by I=$(0,\xi_1)$, II=$(\xi_1,\xi_2)$ and
III=$(\xi_2,\infty)$, where $\xi_1=0.25526$ and $\xi_2=2.6776$. In
the cases I and III the topological squared mass is negative,
while in II is positive. Therefore only the situation II is
consistent. In this case the mean energy and the canonical entropy
are given by
\begin{equation}
E_{II}(\xi)=(\varepsilon^{(r)}+P(\xi))/L
 \label{he24}
\end{equation}
and
\begin{equation}
S_{II}(\xi)=\xi P(\xi)+R(\xi)+cte_{II}.
 \label{he25}
\end{equation}
\subsection{The negative topological squared mass, i.e., $m_T^2<0$}

Now let us consider the case where the values of the compactified
lengths and temperature give us to the situation where the
analytic extension of the homogeneous Epstein zeta function
$Z_4(2,a_1,...,a_4)$ has only positive values. In this case the
topological squared mass is a negative quantity, since we are
considering that the coupling constant is negative $g=-|g|<0$. In
this case we have to impose the second renormalization condition
of the effective potential in a arbitrary point $\phi=M$ different
from zero. If we take $M=0$, the effective potential is not only
not bounded from below, but also will not have any local minimum,
and in this case the system is unstable under small external
perturbations. The second renormalization condition can be written
as
\begin{equation}
\frac{\partial^4 V_{eff}}{\partial\phi^4}\bigg|_{\phi=M}=8\pi^2 g\mu^4.
\label{h23}
\end{equation}
Using the Eq. (\ref{h9}) and Eq. (\ref{h23}) we get the
renormalized effective potential
\begin{eqnarray}
V_{eff}^R(\phi\,;\beta,L_1,L_2,L_3)&=&\mu^2\frac{m_T^2}{2}\phi^2-\frac{\mu^4}
{3}\pi^2|g|\phi^4+U\nonumber\\&&-\frac{1}{\beta\Omega}
\sum_{s=3}^{\infty}\alpha(\phi,s)
Z_4(2s,a_1,a_2,a_3,a_4).
 \label{h24}
\end{eqnarray}
In the Eq. (\ref{h24}) the quantity U is defined in Eq. (\ref{h3}) and $\alpha$ is given by
\begin{equation}
\alpha(\phi,s)=|g|^s\bigg(\frac{\phi^{2s}}{2s}-
\frac{\phi^4}{4!}(2s-1)(2s-2)(2s-3)M^{2s-4}\bigg).
 \label{h25}
\end{equation}
The renormalized effective potential
Eq. (\ref{h24}) can be rewritten in the following way
\begin{equation}
V_{eff}^R(\phi\,;\beta,L_1,L_2,L_3)=-\frac{\mu^2m_T^2}{2}F(\phi;\beta,L_1,L_2,L_3)+U,
\label{h31}
\end{equation}
where we have defined the function $F(\phi;\beta,L_1,L_2,L_3)$ as
\begin{equation}
F(\phi\,;\beta,L_1,L_2,L_3)=-\phi^2+A\phi^4-\sum_{s=3}^{\infty}C_s\phi^{2s}.
\label{h32}
\end{equation}
The coefficients $C_s$, independent of the field $\phi$, are
defined, for $s=3,4,...$, by
\begin{equation}
C_s=\frac{|g|^{s-1}}{s}\frac{Z_4(2s,a)}{Z_4(2,a)},
 \label{h33}
 \end{equation}
and the coefficient $A$ is defined by the expression
\begin{equation}
A=A_o+\frac{1}{4!}\sum_{s=3}^{\infty}C_s(2s)(2s-1)(2s-2)(2s-3)M^{2s-4}
\label{h34}
\end{equation}
where
\begin{equation}
A_o=-\frac{2\beta\Omega\mu^4\pi^2}{3Z_4(2,a)}.
 \label{h35}
\end{equation}
We have denoted for simplicity
$Z_4(2s,a)=Z_4(2s,a_1,a_2,a_3,a_4)$. Note that the coefficients
$C_s$ are defined in the domain of convergence of $Z_4(2s,a)$,
i.e., $s=3,4,...$, therefore we have that  $Z_4(2s,a)>0$ and as we
are considering the case where $Z_4(2,a)>0$, the coefficients
$C_s$ are positive. If we take the second renormalization
condition in a point $M=0$, the coefficient of the fourth power of
the field in Eq. (\ref{h32}) would be negative. In this case is
not possible to find a local minimum of the effective potential.
One way to circumvented this situation is to choose $M$ where the
coefficient $A$ assume a positive value.
\begin{figure}
 \centering
 \includegraphics{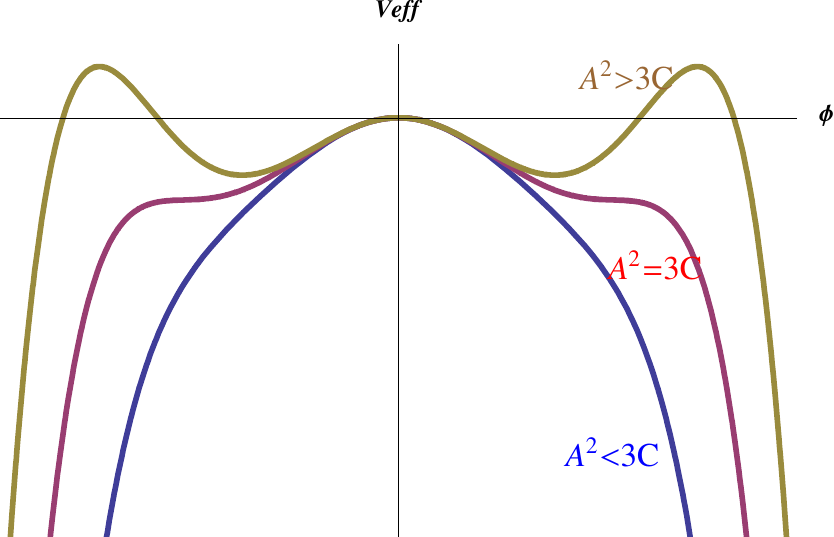}
 \caption{Behavior of the effective potential in the case $m_T^2<0$ for differents values of $A$.}
\label{figure3}
\end{figure}
In the Fig. (\ref{figure3}) the behavior of the effective
potential for different values of $M$, and consequently, for
different values of $A$, is presented for small values of the
field $\phi$ and of the coupling constant. This behavior depends on the first terms of Eq.
(\ref{h32}). In this approximation we are taking into account only
the three first terms in Eq. (\ref{h32}), where we are denoting
the third coefficient $C=C_3>0$, i.e., we are taking
\begin{equation}
F(\phi\,;\beta,L_1,L_2,L_3)=-\phi^2+A\phi^4-C\phi^{6}.
\label{aprox1}
\end{equation}
From Fig. (\ref{figure3}) we have that the only situation where
the effective potential has a local minimum and the theory is
metastable, is taking $M$ where $A$ is positive and $A^2>3C$. This
case is the only one where we can find a local minimum of the
effective potential when the topological squared mass satisfies
the inequality $m_T^2<0$. This minimum is localized outside the
origin and the system develops a condensate. We conclude that we
have to take $M$ in such that $A>\sqrt{3C_3}>0$. In terms of $M$
this inequality can be written as
\begin{equation}
-\frac{2\beta\Omega\mu^4\pi^2}{3Z_4(2,a)}+\frac{1}{4!}
\sum_{s=3}^{\infty}C_s(2s)(2s-1)(2s-2)(2s-3)M^{2s-4}
>\sqrt{\frac{|g|^2 Z_4(6,a)}{Z_4(2,a)}}.
 \label{h36}
\end{equation}
%

\begin{figure}
 \centering
 \includegraphics[scale=0.8]{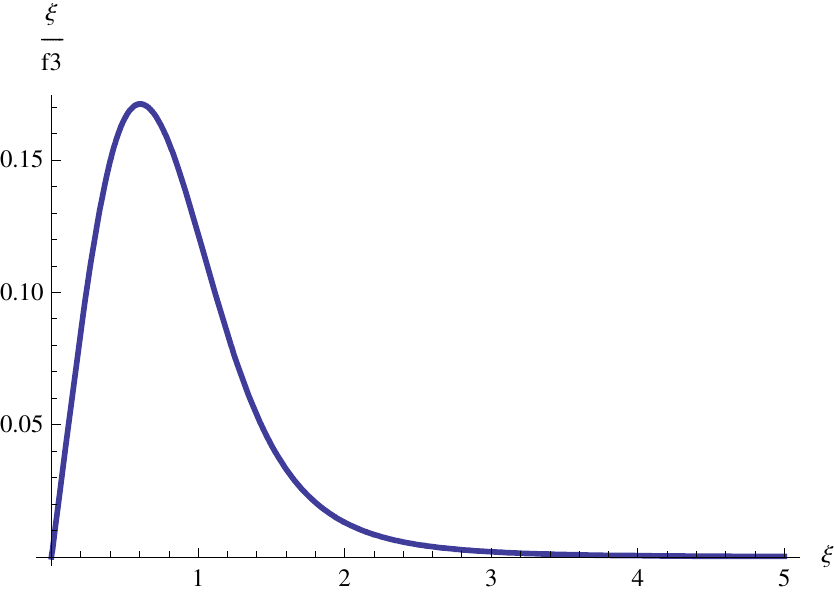} \vspace{1cm}
 \includegraphics[scale=0.8]{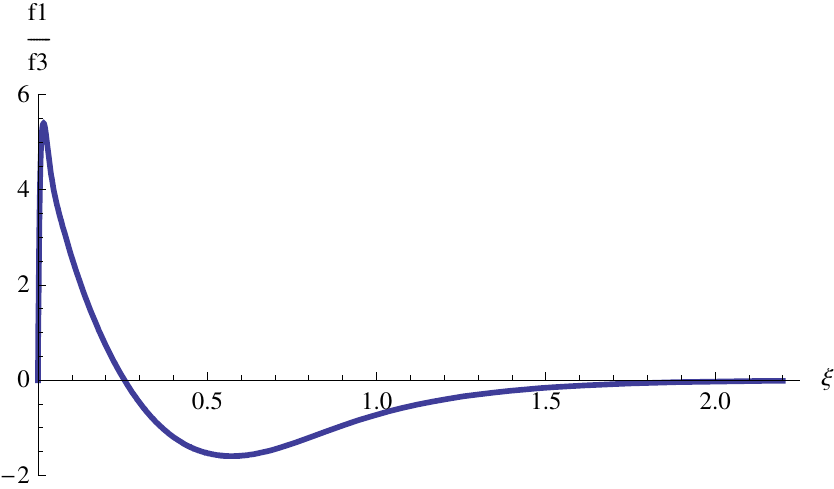}
\caption{The functions $\xi/f_3(\xi)$ and $f_1(\xi)/f_3(\xi)$}
\label{figure4}
\end{figure}

%
%
%
We will show later that, for a given coupling constant and volume
of the compact domain, we can always find $M$ that satisfies Eq.
(\ref{h36}) for any temperature. We can make an approximation in
the series given by Eq. (\ref{h34}) taking only the term $s=3$.
The coefficient $A$ would be
\begin{equation}
A=A_o+15CM^2. \label{h37}
\end{equation}
From now we will consider that the lengths of our compact domain
are the same, $L_1=L_2=L_3=L$, and we will define $\xi=\beta/L$.
It is easy to show that
\begin{equation}
Z_4(2s,a)=(\mu L)^{2s}f_s(\xi)
\label{h38}
\end{equation}
where the function $f_s(\xi)$ is defined by
\begin{equation}
f_s(\xi)=Z_4(2s,1,1,1,\xi^{-1}).
\label{h39}
\end{equation}
Considering Eq. (\ref{h37}) and Eq. (\ref{h38}) the condition Eq.
(\ref{h36}) can be rewritten in the following way
\begin{equation}
M^2(\mu
L)^2>\frac{2\pi^2}{15|g|^2}\frac{\xi}{f_3(\xi)}+
\frac{1}{5|g|}\sqrt{\frac{f_1(\xi)}{f_3(\xi)}}.
\label{h40}
\end{equation}
From Fig. (\ref{figure4}) we see that the functions $\xi/f_3(\xi)$
and $f_1(\xi)/f_3(\xi)$ are bounded from above and then we always
can find a value of $M$ that satisfies Eq. (\ref{h40}). Note that
since we are considering negative topological squared mass we are
taking values of $\xi$ such that $f_1(\xi)$ is positive and, as
$f_3(\xi)$ is always positive. Therefore we are able to take the
square root of $f_1(\xi)/f_3(\xi)$ in the domain where we are
working now. Defining  $v_1$ and $v_2$ as upper bounds of the
functions $\xi/f_3(\xi)$ and $\sqrt{f_1(\xi)/f_3(\xi)}$,
respectively, Eq. (\ref{h40}) can be satisfied by taking
\begin{equation}
M^2(\mu L)^2=\frac{2\pi^2v_1}{15|g|^2}+\frac{v_2}{5|g|}.
\label{h41}
\end{equation}
%
%
%

%
%
%
Using Eq. (\ref{aprox1}) and Eq. (\ref{h37}), we can find the
local minimum of the renormalized effective potential $\phi_0$
given by
\begin{equation}
\phi_0^2=\frac{A-\sqrt{A^2-3C}}{3C}. \label{h42}
\end{equation}
It is better to define $\Theta=\phi_0^2(\mu L)^2$. We have
\begin{eqnarray}
\Theta(\xi)&&=\frac{2\pi^2v_1}{3|g|^2}+\frac{v_2}{|g|}-\frac{2\pi^2}{3|g|^2}\frac{\xi}{f_3(\xi)}
\nonumber\\&&-\frac{1}{3}\Bigg(\frac{4\pi^2}{|g|^2}\Big(\frac{\xi}{f_3(\xi)}\Big)^2-
\frac{12\pi^2}{|g|^2}\Big(\frac{2\pi^2v_1}{15|g|^2}+\frac{v_2}{5|g|}\Big)
\frac{\xi}{f_3(\xi)}+\Big(\frac{2\pi^2v_1}{|g|^2}+\frac{3v_2}{|g|}\Big)
^2-\frac{9}{|g|^2}\frac{f_1(\xi)}{f_3(\xi)}\Bigg)^{1/2}.
\label{h43}
\end{eqnarray}
Considering the parameters $v_1$, $v_2$ and $g$ as constants let
us analyzed the behavior of $\Theta$ with respect to $\xi$.
Performing the Legendre transform in the metastable stationary point
of the renormalized effective potential we get
\begin{eqnarray}
\ln Z(\beta,\Omega)&&=\ln
Z(\beta,\Omega,h)|_{h=0}\nonumber\\&&=-(\beta\Omega)
V_{eff}^R(\phi\,;\beta,L_1,L_2,L_3)|_{\phi=\phi_0}.
 \label{h44}
\end{eqnarray}
Substituting the  Eqs. (\ref{h3}), Eq. (\ref{h31}), Eq.
(\ref{aprox1}), Eq. (\ref{h37}) and Eq. (\ref{h41}) in Eq.
(\ref{h44}) we have
\begin{eqnarray}
\ln Z(\xi)&=&\frac{|g|}{2}f_1(\xi)\Theta(\xi)+\frac{|g|}{2}
\Big\{\frac{2\pi^2\xi}{3}-\Big(\frac{2\pi^2v_1}{3}+|g|v_2\Big)f_3(\xi)\Big\}\Theta^2(\xi)
\nonumber\\&&+ \frac{|g|^4}{3}f_3(\xi)\Theta^3(\xi)-
\sum_{n_1,...,n_3=-\infty}^{\infty\hspace{0.18cm},}
\left(\tilde{n}\pi\xi+\ln\left(1-e^{-2\tilde{n}\pi\xi}
\right)\right)-J_1.
  \label{h45}
\end{eqnarray}
Using Eq. (\ref{imp}), Eq. (\ref{imp1}) and Eq. (\ref{h45}) we obtain
the mean energy
\begin{equation}
E(\xi)=(\varepsilon^{(r)}+P(\xi)+\chi(\xi))/L \label{h46}
\end{equation}
and the canonical entropy
\begin{equation}
S(\xi)=\xi P(\xi)+R(\xi)+\psi(\xi)+cte
\label{h47}
\end{equation}
where the functions $\varepsilon^{(r)}$, $P(\xi)$ and $R(\xi)$ are
defined in Eq. (\ref{er}), Eq. (\ref{P}) and Eq. (\ref{R}),
respectively. The functions $\chi(\xi)$ and $\psi(\xi)$ are given
by the expressions
\begin{equation}
\chi(\xi)=
-\frac{|g|}{2}\Big\{f'_1(\xi)\Theta(\xi)+\Big(\frac{2\pi^2}{3}-
\Big(\frac{2\pi v_1}{3}+|g|v_2\Big)f'_3(\xi)\Big)\Theta^2(\xi)
+\frac{|g|^2}{3}f'_3(\xi)\Theta^3(\xi)\Big\}
\label{h48}
\end{equation}
and
\begin{equation}
\psi(\xi)= \frac{|g|}{2}\Big\{g_1(\xi)\Theta(\xi)+\Big(\frac{2\pi
v_1}{3}+ |g|v_2\Big)g_3(\xi)\Theta^2(\xi)
+\frac{|g|^2}{3}g_3(\xi)\Theta^3(\xi)\Big\}. \label{h49}
\end{equation}
Since we are considering here that the topological squared mass is
negative, these results are valid only in the intervals
I=$(0,\xi_1)$ and III=$(\xi_2,\infty)$ of the variable $\xi$. This
results can be expressed in the following way. We have
\begin{equation}
E_{(I,III)}(\xi)=(\varepsilon^{(r)}+P(\xi)+\chi(\xi))/L
\label{h50}
\end{equation}
and
\begin{equation}
S_{(I,III)}(\xi)=\xi P(\xi)+R(\xi)+\psi(\xi)+cte_{(I,III)}.
\label{h51}
\end{equation}
In Eq. (\ref{h50}) and Eq. (\ref{h51}) we see explicitly that the
form of the mean energy is the same in the regions I and III, but
the form of the canonical entropy is different in each of these
intervals. This discrepancy is due to certain constants, $cte_I$
and $cte_{III}$, that will be fixed with the help of the third law
of thermodynamics and assuming the continuity of the entropy.

\section{Analysis of the results}
\begin{figure}
 \includegraphics[width=17cm,height=5.5cm]{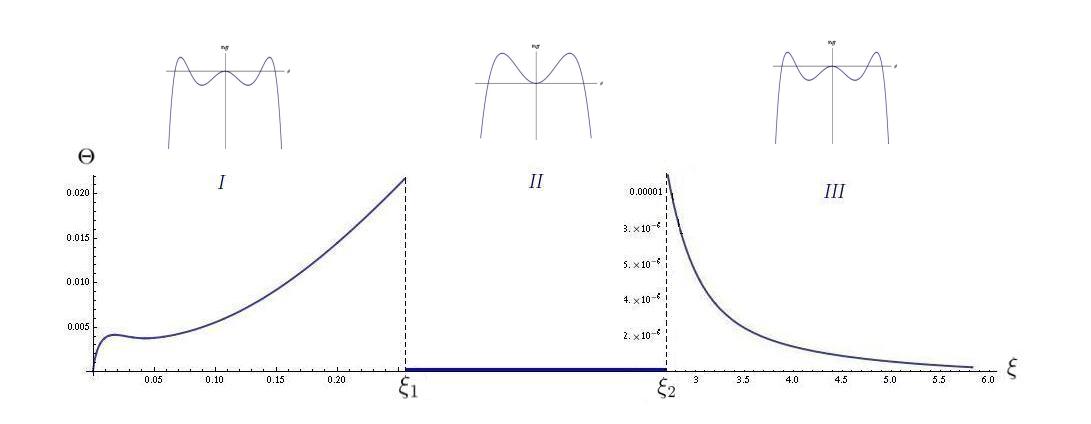}
\caption{The minimum of the effective potential $\Theta(\xi)$ and the form of the effective potential for
different values of $\xi$.}
\label{figure5}
\end{figure}

We have found that due to the boundary conditions imposed on the
field and the presence of a thermal reservoir, there is a
topological and thermal generation on mass. This topological mass
depends on the lengths of the compactification of the spatial
coordinates and on the temperature. It was shown that the sign of
the topological squared mass is crucial to determine the profile
of the effective potential. Then we obtained two different
physical situations: the case where the topological squared mass
is positive and the case where it is negative. We shown that when
the topological squared mass is negative the system develops a
condensate. In this case, the minimum of the effective potential
is not localized at the origin and it is given by the function
$\Theta(\xi)$ defined in Eq. (\ref{h43}). We would like to stress
that only in the intervals I and III of the variable $\xi$ the
topological squared mass is negative. In the interval II of $\xi$
the topological squared mass is positive and the effective
potential has a trivial minimum. The Fig. (\ref{figure5}) shows
the minimum of the effective potential, $\Theta$, as a function of
$\xi$, for the values $v_1=100$, $v_2=100$ and $|g|=0.13$. There
also is presented the form of the effective potential in each of
the three ranges of values of $\xi$. From  Fig. (\ref{figure5}) we
see that the minimum of the effective potential is at the origin
when we are considering very high temperatures, $\xi\rightarrow
0$, or when we are considering very low temperature,
$\xi\rightarrow\infty$. From this last result we have that the
function $\psi(\xi)$ goes to zero when the temperature tends to
zero.
\\
We have found
the entropy formulas in each intervals of values of $\xi$ up to certain constants
\begin{eqnarray}
&&S_{I}(\xi)=\xi P(\xi)+R(\xi)+\psi(\xi)+cte_{I},
\nonumber\\&& S_{II}(\xi)=\xi P(\xi)+R(\xi)+cte_{II},
\nonumber\\&& S_{III}(\xi)=\xi P(\xi)+R(\xi)+\psi(\xi)+cte_{III}.
\label{h52}
\end{eqnarray}
Using the third law of thermodynamics $\lim_{\xi\rightarrow\infty}S_{III}=0$,
assuming the continuity of the entropy with the parameter $\xi$:
$S_{I}(\xi_1)=S_{II}(\xi_1)$ and $S_{II}(\xi_2)=S_{III}(\xi_2)$,
and using the fact that the functions $P(\xi)$, $R(\xi)$ and $\psi(\xi)$ go to zero
when $\xi\rightarrow\infty$, we can fix the constants that appear in the formulas of the entropies
\begin{eqnarray}
&& cte_{I}=\psi(\xi_2)-\psi(\xi_1),
\nonumber\\&&cte_{II}=\psi(\xi_2), \nonumber\\&&cte_{III}=0.
\label{h53}
\end{eqnarray}
For generic values of the parameters $(v_1,v_2,g)$, the function
$\psi(\xi)$ is not positive defined and the entropy can be
negative for some values of $\xi$. For large values of $v_1$ and
$v_2$ and small $g$ this situation is excluded.
\\\\
With the thermodynamics quantities, the validity of the Bekenstein
bound can be verified for the system. The Bekenstein bound states
that $S/E\leq2\pi R$, where $R$ is the smallest ratio of the
sphere that circumscribe our finite spatial domain. Since we are
considering that all our compactified lengths are equals to $L$,
we have that $R=\sqrt{3}L/2$. Defining the function $T=S/2\pi RE$
on each of the intervals I, II and III and using Eq. (\ref{he24}),
Eq. (\ref{h50}), Eq. (\ref{h52}) and Eq. (\ref{h53}) we have that
\begin{figure}[t]
 \centering
 \includegraphics{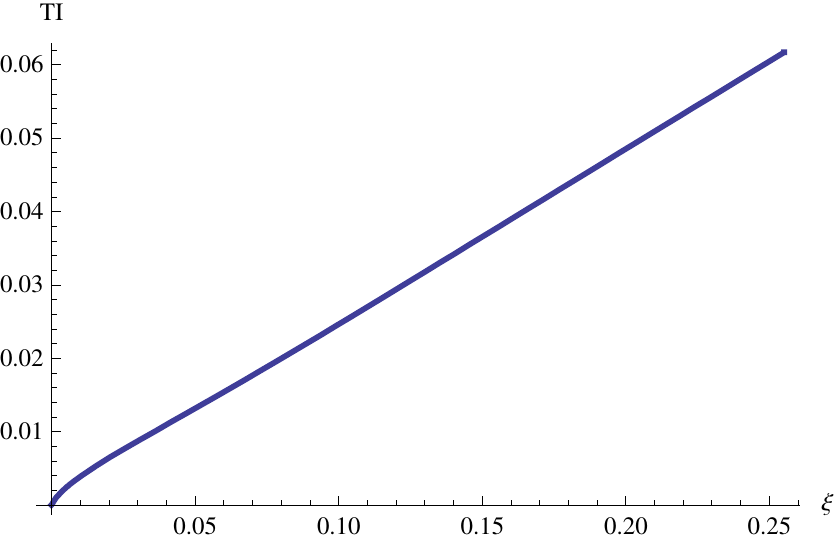}
 \caption{The function $T_I(\xi)$ in its domain $\xi\in$ I=$(0,\xi_1)$}
 \label{figure6}
\end{figure}
\begin{equation}
T_I(\xi)=\frac{1}{\sqrt{3}\pi}\frac{\xi
P(\xi)+R(\xi)+\psi(\xi)+\psi(\xi_2)-\psi(\xi_1)}
{\varepsilon^{(r)}+P(\xi)+\chi(\xi)}
\label{h54},
\end{equation}
\begin{equation}
T_{II}(\xi)=\frac{1}{\sqrt{3}\pi}\frac{\xi
P(\xi)+R(\xi)+\psi(\xi_2)}{\varepsilon^{(r)}+P(\xi)} \label{h55},
\end{equation}
\begin{equation}
T_{III}(\xi)=\frac{1}{\sqrt{3}\pi}\frac{\xi
P(\xi)+R(\xi)+\psi(\xi)}{\varepsilon^{(r)}+P(\xi)+\chi(\xi)}
\label{h56}.
\end{equation}
Each of these functions are valid only when $\xi$ is defined in
the domains I, II and III respectively.
\begin{figure}[h]
 \centering
 \includegraphics{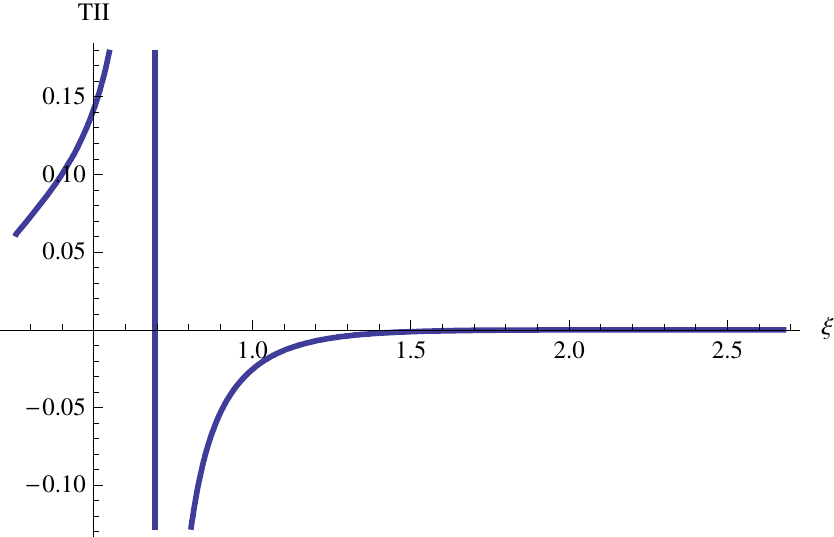}
 \caption{The function $T_{II}(\xi)$ in its domain $\xi\in$ II=$(\xi_1,\xi_2)$}
 \label{figure7}
\end{figure}
In Fig. (\ref{figure6}) we have the function $T_I(\xi)$ for
$\xi\in$ I=$(0,\xi_1)$, there we have used the values $v_1=100$,
$v_2=100$ and $|g|=0.13$. In this situation we have that the field
exhibits a condensate. In this regime of high temperatures, we
expected that the negative Casimir energy of the system would be
irrelevant to the Bekenstein bound, since as we can be see in Fig.
(\ref{figure6}), the thermal fluctuations dominates over any
quantum contributions and the Bekenstein bound is satisfied in
this situation.
\\

In Fig. (\ref{figure7}) we have the function $T_{II}(\xi)$ in the
region $\xi\in$ II=$(\xi_1,\xi_2)$. In this regime the
renormalized effective potential has a trivial minimum and the
system behaves as a free bosonic gas. Since we are considering a
compact domain with periodic boundary conditions on the spatial
coordinates we have that the renormalized Cassimir energy is
negative, $\varepsilon^{(r)}=-0.81$. From Fig. (\ref{figure7}) we
see that from some value $\xi'$, defined by the equation
$\varepsilon^{(r)}+P(\xi')=1$, the function $T_{II}(\xi)$ begins
to take values greater than one and the Bekenstein bound is
violated. It was found that $\xi'=0.6720$. In Fig. (\ref{figure7})
also can be seen a divergent point $\xi_d$ given by
$\varepsilon^{(r)}+P(\xi_d)=0$. Since the sign of the Casimir
energy is negative, the Bekenstein bound is violated.
\\
\begin{figure}[h]
 \centering
 \includegraphics[scale=0.85]{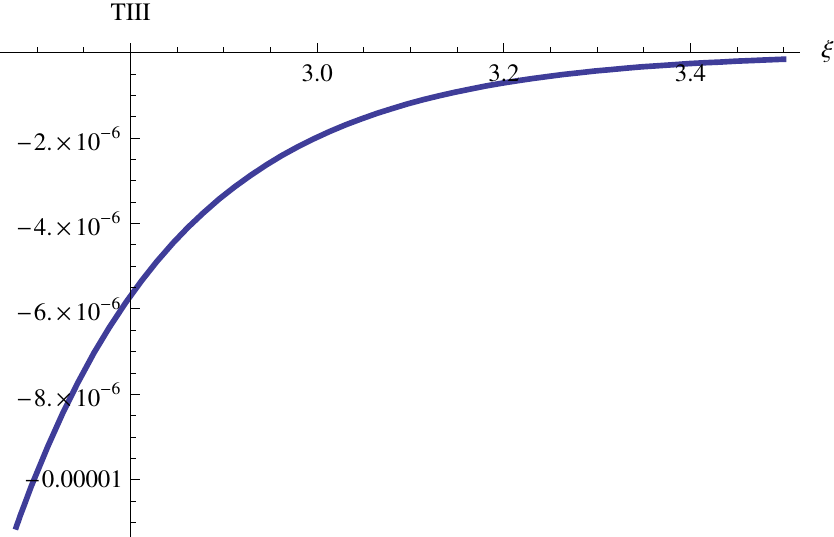} \vspace{1cm}
 \includegraphics{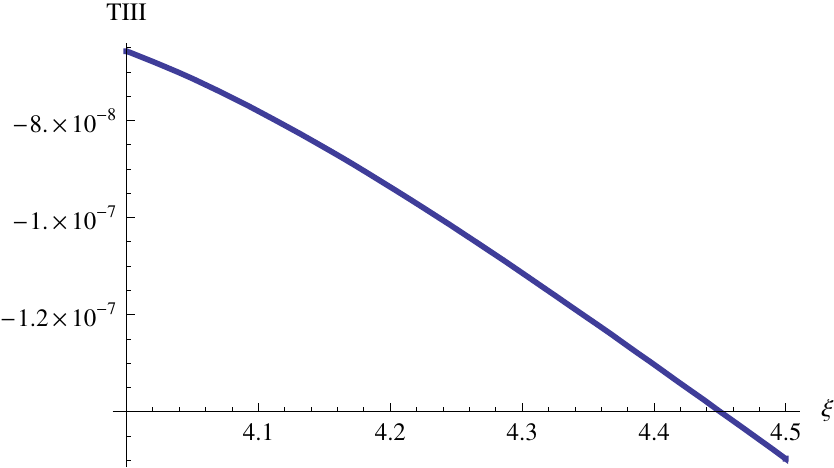}
 \caption{The function $T_{III}(\xi)$ in its domain $\xi\in$ III=$(\xi_2,\infty)$}
 \label{figure8}
\end{figure}

In the domain III our theory also exhibits a condensate. Since in
this regime we are considering low temperatures, the quantum
fluctuations dominate over the thermal one. The Fig.
(\ref{figure8}) shows that $T_{III}(\xi)$ is negative, this is
because the negative Casimir prevails over the condensate
contribution making the total mean energy of the system negative.
Since the entropy is always positive, the Bekenstein bound is also
violated in this situation.

Then we shown that there is an intrinsically information storage
capacity limit for the $(\lambda\,\varphi^{\,4})_{d}$ field theory
with the negative sign of the coupling constant, for values of the
temperature greater that certain critical temperature given by
$T_{cr}=1/L\xi'$. For temperatures lower than $T_{cr}$ the
Bekenstein bound in invalidated mainly due to the negative Casimir
energy. The asymptotically freedom of the model and the presence
of the condensate do not change the discussion about the quantum
bound. In conclusion the main feature in the discussion of the
validity of the Bekenstein bound is related to the sign of the
zero point energy of the system.

\section{Acknowlegements}
This paper was supported by Conselho Nacional de Desenvolvimento
Cientifico e Tecnol{\'o}gico do Brazil (CNPq), Funda\c{c}\~ao de
Amparo a Pesquisa do Rio de Janeiro (FAPERJ) and Funda\c{c}\~ao de
Amparo a Pesquisa de S\~ao Paulo (FAPESP). We also want to thank to the
LAFEX neutrino group (CBPF) for the support with its OLYMPUS computer cluster.

\end{document}